\documentclass[twocolumn,english,aps,prl,superscriptaddress,longbibliography]{revtex4-1}
\usepackage[T1]{fontenc}
\usepackage[latin9]{inputenc}
\usepackage{amsmath}
\usepackage{amssymb}
\usepackage{mathdots}
\usepackage{graphicx}

\makeatletter
\usepackage{epsfig}
\usepackage{dcolumn}
\usepackage{bm}
\usepackage{bbm}
\usepackage{amsfonts}
\usepackage{latexsym}
\usepackage{color}
\usepackage[normalem]{ulem}

\makeatother

\usepackage{babel}
\begin{document}

\title{Topological domain wall states in a non-symmorphic chiral chain}

\author{Wojciech Brzezicki}

\affiliation{International Research Centre MagTop, Institute of Physics, Polish
Academy of Sciences,~\\
 Aleja Lotnikow 32/46, PL-02668 Warsaw, Poland}

\author{Timo Hyart}

\affiliation{International Research Centre MagTop, Institute of Physics, Polish
Academy of Sciences,~\\
 Aleja Lotnikow 32/46, PL-02668 Warsaw, Poland}
\begin{abstract}
The Su-Schrieffer-Heeger (SSH) model, containing dimerized hopping
and a constant onsite energy, has become a paradigmatic model for one-dimensional topological phases, soliton excitations and fractionalized charge  in the presence of chiral symmetry. Motivated by the recent developments in engineering artificial lattices, we study an alternative model
where hopping is constant but the onsite energy is dimerized. We find that it has a non-symmorphic
chiral symmetry and supports topologically distinct phases described by a $\mathbb{Z}_{2}$ invariant $\nu$. In the case of multimode ribbon we also find topological phases protected by hidden
symmetries and  we uncover the corresponding $\mathbb{Z}_{2}$ invariants  $\nu_{n}$. We show that, in contrast to the SSH case, zero-energy states do not necessarily appear at the boundary between topologically distinct phases, but instead these systems support a new kind of bulk-boundary correspondence: The energy of the topological domain wall
states typically scales to zero as $1/w$, where  $w$ is the width of the domain wall separating phases with different topology. Moreover, under specific circumstances we also find a faster scaling $e^{-w/\xi}$, where $\xi$ is an intrinsic length scale.
We show that the spectral flow of these states and  the charge of the domain walls are different than in the case of the SSH model.
\end{abstract}

\maketitle

\begin{figure}[hbt!]
\begin{centering}
\includegraphics[width=0.741\columnwidth]{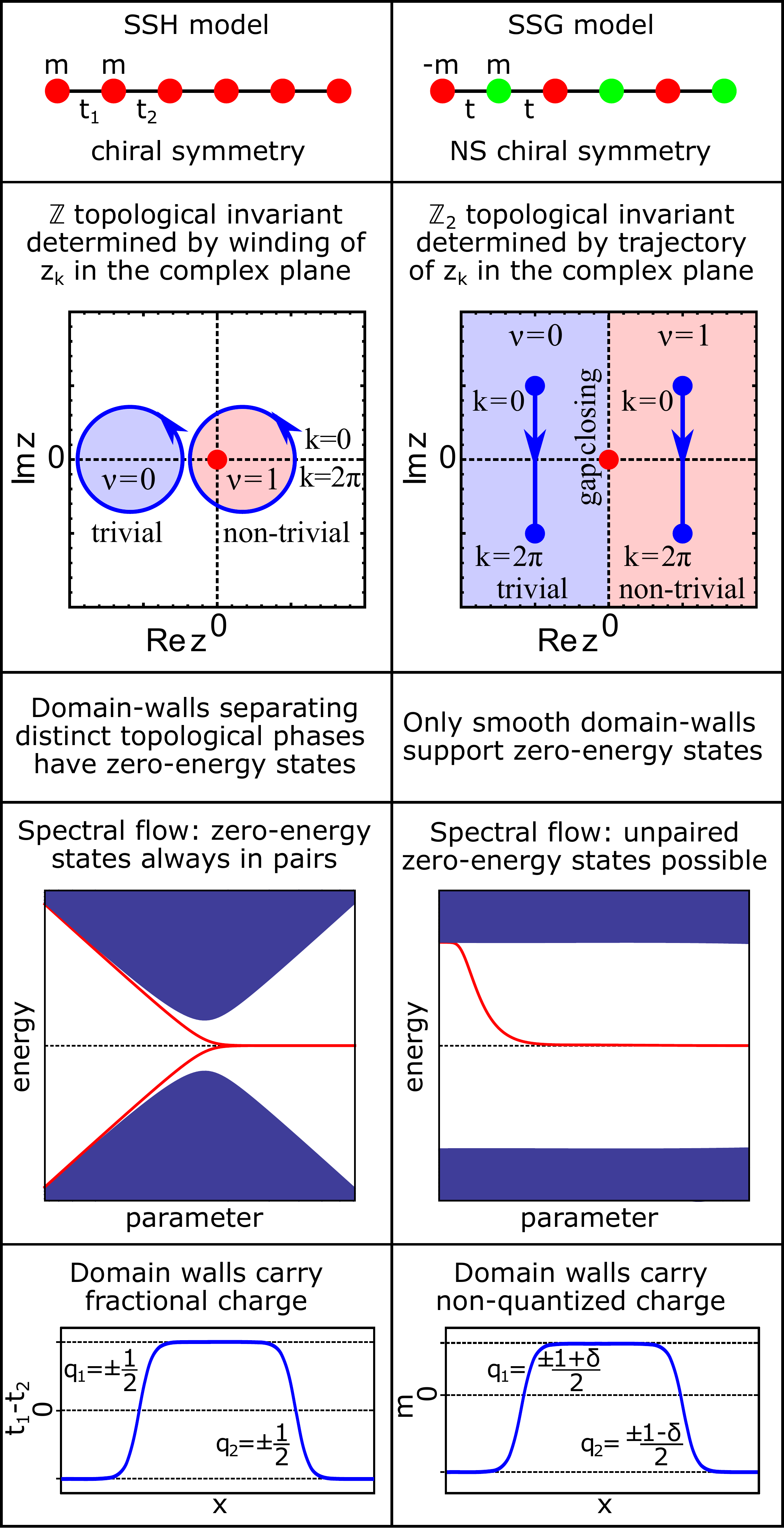} 
\par\end{centering}
\caption{Comparison of SSH and SSG models. The topological invariants are determined by the trajectory of the determinant $z_{k}$ of the off-diagonal block of the Hamiltonian as a function $k$ and the DWs separating topologically distinct phases lead to bound states, but in SSG model the bound state energy depends on the DW width, and the spectral flow and charge of the DW are different than in the SSH model.
}
\vskip -0.8 cm
\label{fig1} 
\end{figure}

The Su-Schrieffer-Heeger (SSH) model was originally introduced to describe the properties of conducting polymers, where the spontaneous symmetry breaking leads to dimerization of the sites along the chain \cite{SSH79,SSH80,SSH88}.  Due to two-fold degeneracy of the ground state  a new type of excitation, a domain wall (DW) between different bonding structures, can exist. For the conducting polymers the width of the DW excitations is large and  they can propagate along the chain. Thus, they can be considered as solitons in analogy to the shape-preserving propagating solutions of the nonlinear differential equations \cite{SSH88}. Moreover, the solitons in the SSH model have a remarkable effect on the electronic spectrum leading to an appearance of a bound state in the middle of the energy gap. This midgap state is understood as a topologically protected boundary mode and the SSH model serves as a paradigmatic example of chiral symmetric topological insulator \cite{Kane2010, Chiu2016}. Namely, the chiral symmetry allows to block-off-diagonalize the Hamiltonian and the winding of the determinant $z_{k}$ of the off-diagonal block around the origin as a function $k$ determines a topological invariant (see Fig.~\ref{fig1}). Because this invariant is different on two sides of the DW, each DW carries zero-energy bound state. The DWs come in pairs so that the spectral flow is symmetric around zero-energy  and each DW carries a charge $q=\pm1/2$ in analogy to the fractionally charged excitations studied in the quantum field theory  \cite{Jakiw76}.
This can be also understood in terms of modern notions of bulk obstructions and filling anomalies \cite{Kha19}.

The idea of soliton excitations reappears in the context
of 1D diatomic polymers in a form of the Rice-Mele model \cite{Rice82,Brazowskii80,Jackiw83}, which has been studied also in contexts of ferroelectricity
\cite{Vaderbilt93,Onoda04} and  organic salts \cite{Soos07,Tsuchiizu16}. In this model
 not only bond length alternates but also the onsite energy (mass) takes
opposite sign for the even/odd lattice sites (Fig. \ref{fig1}).
A very interesting
feature of such a model is that its solitons can carry irrational
charge $q=\frac{1}{2}(1\pm f)$ \cite{Rice82}, where $f$ describes the breaking of the  chiral symmetry \cite{Wilczek81}. The interest for these models has revived because they can engineered in photonic systems \cite{Kitagawa12, Schomerus}, optical lattices \cite{Atala13, Wang13, Lohse16, Nakajima16, Leder16, Meier16} and nanostructures
\cite{Yan19, Park16, Drost17, Huda18, Fischer18, Fasel18}  in a controlled way, and in these systems also their emergence from spontaneous symmetry breaking  \cite{Weber 19} and the properties of the solitons can be tuned using external parameters \cite{Pupillo08, Herrera2011, Hague12, Mezzacapo12, Bissbort13, Marcin18}. Motivated by the new possibilities opened by these recent
developments we focus  on a special case of the Rice-Mele
model where all the hopping amplitudes are equal and only the mass
term alternates. We show that in this case the model has an interesting non-symmorphic (NS) chiral symmetry and it supports a topologically nontrivial phase described by a non-symmorphic chiral $\mathbb{Z}_{2}$ invariant $\nu$. This invariant was found by Shiozaki, Sato and Gomi in their pioneering work on non-symmorphic topological insulators \cite{Sato15} and therefore we name the special case of the Rice-Mele model as the Shiozaki-Sato-Gomi (SSG) model.  The peculiar property of the SSG model is that the bulk topological
invariant does not guarantee the existence of the end states in an open system, because the boundary always breaks the NS chiral symmetry \cite{Sato15}.

\begin{figure}[t]
\begin{centering}
\includegraphics[width=1\columnwidth]{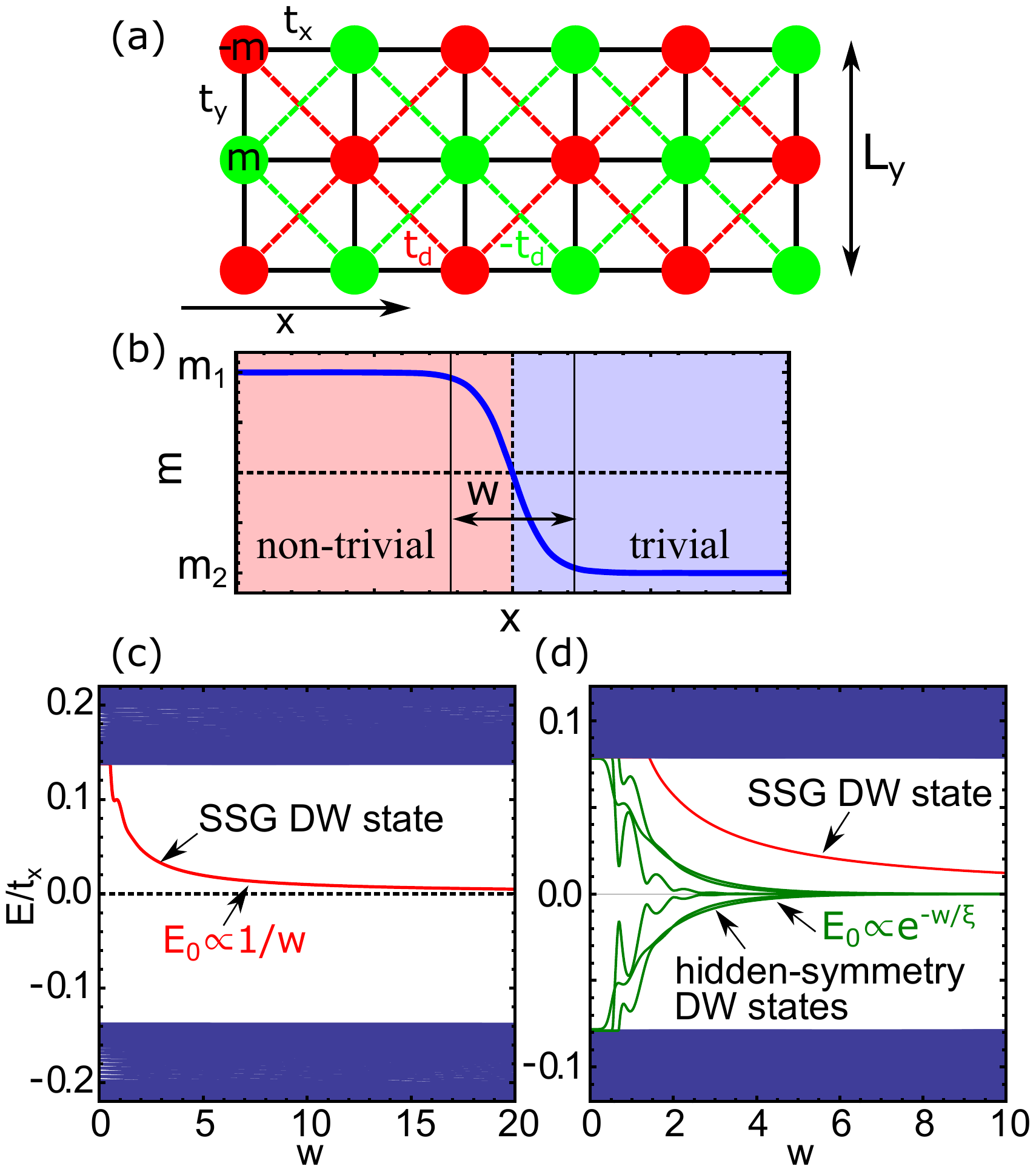} \vspace{-0.6cm}
\par\end{centering}
\caption{(a) Schematic view of the multimode SSG model. (b) Domain wall separating
different mass $m$ regions with different NS chiral $\mathbb{Z}_{2}$ invariants $\nu$ or hidden $\mathbb{Z}_{2}$ invariants $\nu_n$. (c) Spectral flow of the DW states for multimode SSG chain as a function of the DW width $w$ for $L_y=7$, $L_{x}=1000$ and $t_y=t_x$. DW separates regions with masses $m_1=0.2t_x$, $m_2=-20t_x$  and  $t_d=0.6t_x$ differing only by $\nu$. There is a single DW state whose energy approaches zero $\propto 1/w$. (d) The same for DW separating regions differing both by $\nu$ and $\nu_n$ with $m_1=0.25t_x$, $m_2=-10t_x$ and $t_d=0.05t_x$. In addition to the unpaired DW state with energy approaching zero  $\propto 1/w$, there are hidden-symmetry protected DW states with energies approaching zero $\propto e^{-w/\xi}$, where $\xi$ is an intrinsic length scale.
 }
\label{fig2} 
\end{figure}

In this paper we analytically derive
exact phase diagrams of SSG nanoribbons of arbitrary width  and uncover hidden symmetries
relying on interchange of transverse and longitudinal modes. In addition to the NS chiral $\mathbb{Z}_{2}$ invariant $\nu$ the multimode ribbons 
 support $\mathbb{Z}_{2}$ invariants  $\nu_{n}$ protected by the hidden
symmetries.
These invariants lead to a new kind of bulk-boundary correspondence: The energy of the topological domain wall
states typically scales to zero as $1/w$, where  $w$ is the width of the domain wall separating phases with different topology. Moreover, under specific circumstances we also find a faster scaling $e^{-w/\xi}$, where $\xi$ is an intrinsic length scale (Figs.~\ref{fig2}, \ref{fig4} and \ref{fig5}). 
The NS chiral symmetry in SSG model leads to several important differences in comparison to the SSH model (Fig.~\ref{fig1}): (i) In SSH model the topological zero-energy end or DW states come in pairs and have zero energy for any DW width, whereas the SSG model supports unpaired DW states approaching zero energy with increasing $w$.  Note that here we consider finite chains with open boundary conditions and the chains are made out of integer number of unit cells. An inifinite SSH chain could also have a unpaired DW state but in finite system the end states always guarantee the the zero-energy states come in pairs. The SSG model is different because even in this kind of situation we can obtain a single zero-energy state at the smooth domain wall but there is no low-energy states at the sharp interface with vacuum. (ii) In SSH model the charge of the DWs is $q=\pm\frac{1}{2}$, whereas for the SSG model we get  irrational charges $q=\frac{\pm1-\delta}{2}, \frac{\pm1+\delta}{2}$ for solitons and antisolitons depending on whether the zero-energy  state is occupied or empty. The DWs in SSG model separate
regions with different onsite energies $\pm m_{1}$ and $\pm m_{2}$ (mass terms) in the two sublattices (see Figs.~\ref{fig1} and \ref{fig2}), and $\delta=\frac{1}{2}(\zeta_{2}-\zeta_{1})$, where $\zeta_{i}$
is the difference of the bulk filling factors of the two sublattices in the region with mass $m_{i}$.

The $k$-space SSG Hamiltonian for a multimode wire is 
\begin{eqnarray}
{\cal H}_{k} & = & -m\sigma_{z}\tau_{z}+2t_{x}\cos\tfrac{k}{2}(\cos\tfrac{k}{2}\sigma_{x}-\sin\tfrac{k}{2}\sigma_{y})\nonumber \\
 & + & t_{y}\tau_{x}+2t_{d}\cos\tfrac{k}{2}(\sin\tfrac{k}{2}\sigma_{x}+\cos\tfrac{k}{2}\sigma_{y})\tau_{y},\label{eq:hamk}
\end{eqnarray}
where $m$ is the mass, $t_{x}$
and $t_{y}$ are hopping amplitudes in $x$
and $y$ directions and $t_{d}$ is the diagonal hopping amplitude with opposite signs in the two sublattices [Fig. \ref{fig2}(a)]. Here  $\sigma_{\alpha}$ are
Pauli matrices describing  the unit cell in the $x$ direction, $\tau_{\alpha}$ are $L_{y}\times L_{y}$ matrices describing
transverse hopping $\left(\tau_{x}\right)_{pq}=\delta_{1,|p-q|}$, $\left(\tau_{y}\right)_{pq}=i\delta_{1,|p-q|}\left(-1\right)^{p}$,
$\left(\tau_{z}\right)_{pq}=\delta_{pq}\left(-1\right){}^{p}$ and $L_{y}$ is the width in the $y$ direction.  For $L_{y}=2$
operators $\tau_{\alpha}$ are equivalent to $\sigma_{\alpha}$ and
in the special case $L_{y}=1$ we set $\tau_{z}=1$
and $\tau_{x}=\tau_{y}=0$. Because the SSG model belongs to symmetry class AI \cite{Chiu2016} (for list of symmetries see
Appendix \ref{sec:Symmetries}) the only known topological invariant is the NS chiral $\mathbb{Z}_{2}$ invariant $\nu$ \cite{Sato15}.

To calculate the topological invariant $\nu$ we rewrite the Hamiltonian in a block off-diagonal form in
the eigenbasis of NS chiral operator $S_{k}=\sin\tfrac{k}{2}\sigma_{x}\tau_{z}+\cos\tfrac{k}{2}\sigma_{y}\tau_{z}$
that anticommutes with ${\cal H}_{k}$. The determinant $z_{k}$ of
the off-diagonal block is a complex number and its trajectory in the
complex plane as a function $k$ determines the topological invariant \cite{Sato15}. Namely, due to non-symmorphicity
of $S_{k}$ the period of $z_{k}$ is $4\pi$ and in a properly chosen basis it satisfies  constraints $\mathfrak{Im}z_{k}=-\mathfrak{Im}z_{k+2\pi}$
and $\mathfrak{Re}z_{k}=\mathfrak{Re}z_{k+2\pi}$, so that the trajectory $z_{k}$ starts at $k=0$ and ends at $k=2\pi$ with same real part but with opposite imaginary part. Thus the parity of the number of times
the trajectory $z_{k}$ crosses the positive real semiaxis for $k\in[0,2\pi]$
is a $\mathbb{Z}_{2}$ topological invariant because it cannot be
changed without closing the gap or breaking the NS chiral symmetry (see Fig.~\ref{fig1}). 
In our case the mirror symmetry $M_{x}=\cos\tfrac{k}{2}-i\sin\tfrac{k}{2}\sigma_{z}$
becomes identity in the eigenbasis of $S_{k}$ (see Appendix  \ref{sec:Symmetries}) so that 
$z_{k}=z_{-k}$. For this reason $\mathfrak{Im}z_{\pi}=0$ and the
formula for the $\nu$ gets simplified to 
\begin{equation}
\nu={\rm sign\,}\mathfrak{Re}z_{\pi}
\end{equation}
in analogy to the simplification of the invariant for topological insulators in the presence of inversion symmetry \cite{Fu07}. The band-inversion
corresponding to a change of $\nu$ happens at $k=\pi$ and $m=0$. 
We find that
\begin{equation}
\nu=\begin{cases} \tfrac{1}{2}[1+{\rm sign}(m)] &  {\rm if } \ L_{y}=2n-1 \\
			0, &  {\rm if } \  L_{y}=2n  \ (n\in\mathbb{N}_{+})
 \end{cases}
\end{equation}

\begin{figure}[t]
\begin{centering}
\includegraphics[width=1\columnwidth]{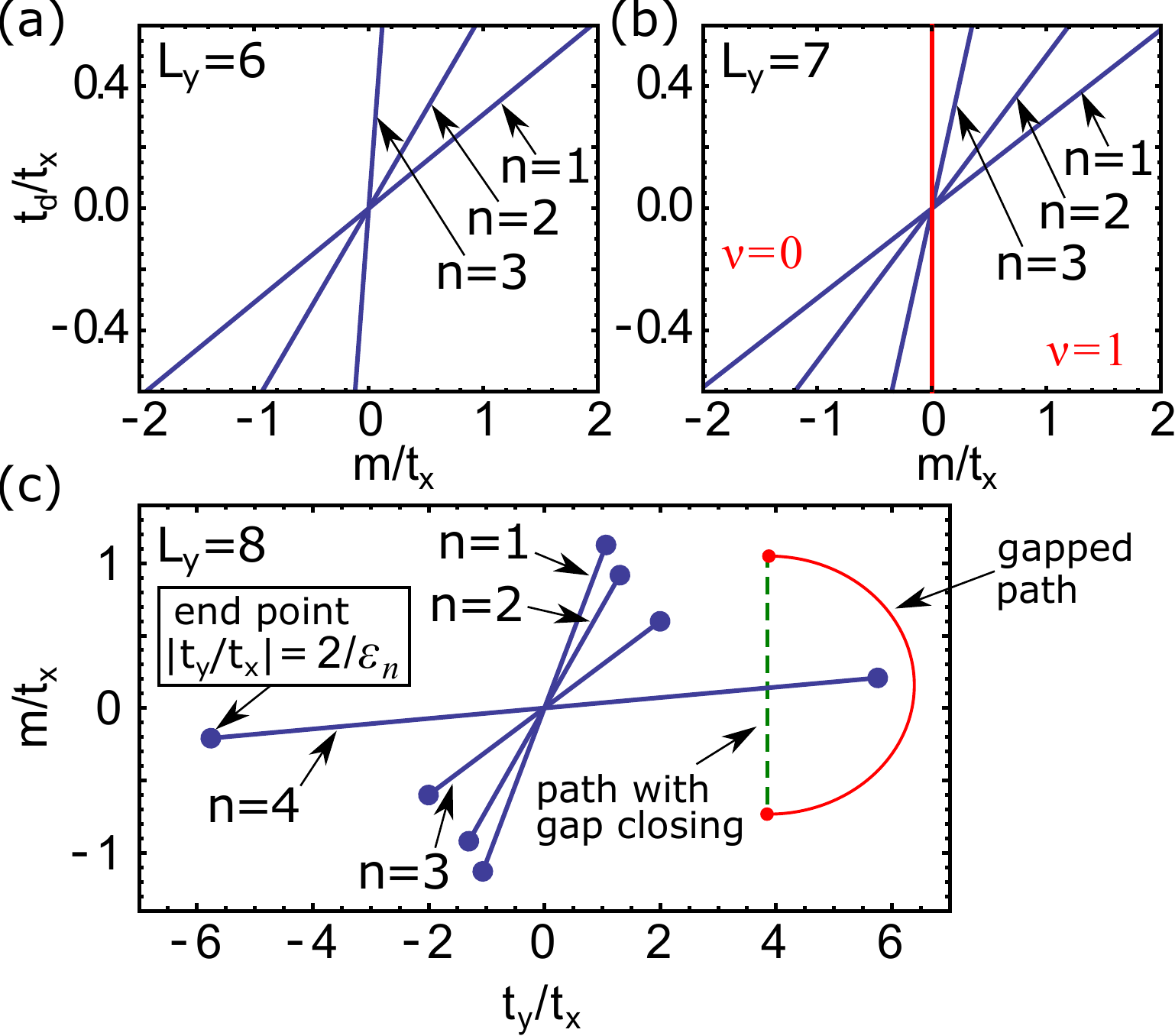} \vspace{-0.6cm}
\par\end{centering}
\caption{(a), (b) Phase diagrams of the multimode SSG model in the $m$-$t_{d}$
plane for $L_y=6,7$ and $t_y=t_x$. The phase transition line between different $\nu$ phases protected by the NS chiral symmetry is shown in red and  phase transition lines between different $\nu_n$ protected by hidden symmetries are shown in blue. (c) Phase diagram in the $m-t_{y}$ plane for $L_y=8$ and $t_d=0.3t_x$.  Each hidden symmetry $n=1,2,\dots,\left\lfloor L_{y}/2\right\rfloor$ exists only if condition (\ref{hiddensymmetrycondition}) is satisfied. Therefore, it is possible to connect phases with different $\nu_n$ without closing the bulk gap by choosing a path in the parameter space which goes outside the region where the hidden symmetry exists.}
\label{fig3} 
\end{figure}

\begin{figure}[t]
\begin{centering}
\includegraphics[width=1\columnwidth]{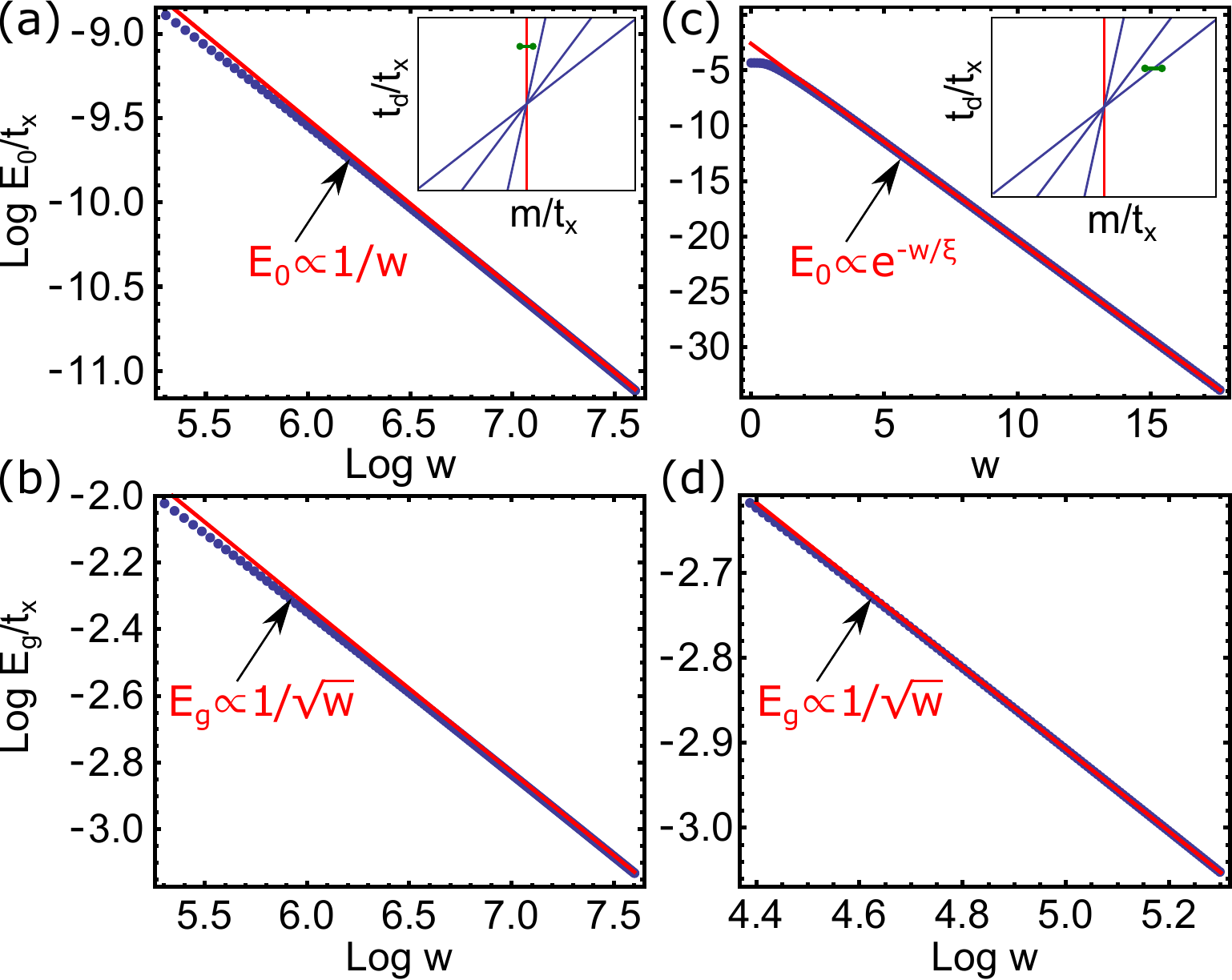} \vspace{-0.6cm}
\par\end{centering}
\caption{Scaling of the energy of the topological DW state $E_{0}$ and the lowest energy of the non-topological states $E_g$ for a multimode SSG model as a function of the DW width $w$. (a), (b)
DW between phases with $m_1=0.12t_x$ and $m_2=-0.12t_x$  distinguished by NS chiral invariant $\nu$. Here we have chosen $t_d=0.4t_x$. (c), (d) DW between phases with $m_1=0.7t_x$ and $m_2=0.1t_x$  distinguished by the hidden symmetry protected invariant $\nu_n$. Here we have chosen $t_d=0.25t_x$. Insets show the phases which are separated by the DWs. The other parameters are $t_y=t_x$, $L_{y}=7$ and  $L_{x}=10000$.}
\label{fig4} 
\end{figure}

\begin{figure}[t]
\begin{centering}
\includegraphics[width=1\columnwidth]{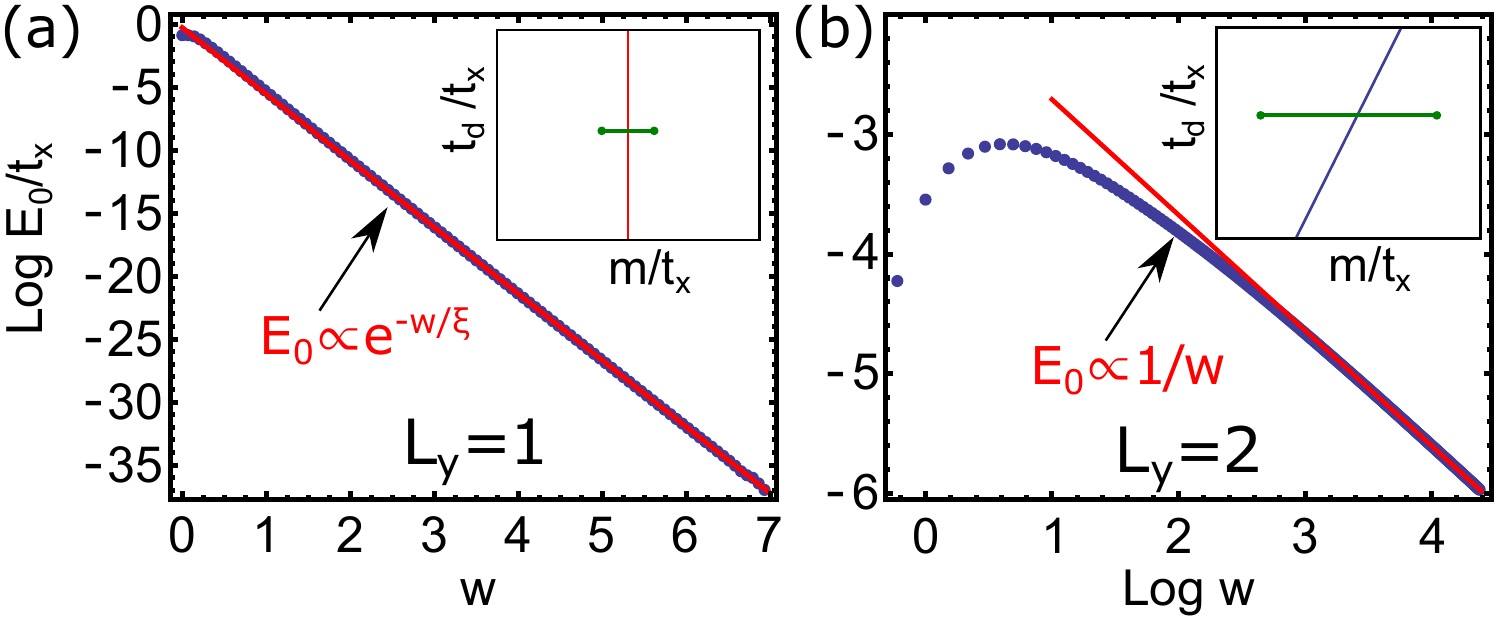} \vspace{-0.6cm}
\par\end{centering}
\caption{Dependence of the scaling of the energy $E_0$  on the microscopic details of the DW. (a) $E_0$ for a DW between phases distinguished by $\nu$-invariant can show exponential scaling if the masses in the two sublattices vary as $\tanh\tfrac{2 i-1}{2w}$ and $-\tanh\tfrac{2i}{2w}$, where $i$ is a position of the unit cell along the chain. Notice that in our default DWs, where the $1/w$ scaling is obtained, the mass terms are a given by $\pm\tanh\tfrac{i}{w}$. The other parameters are $m_{1,2}=\pm t_x$ and $L_y=1$.
(b) $E_0$ for a DW between phases distinguished by  $\nu_n$-invariant can show $1/w$ scaling if the masses vary as $\pm\tanh\tfrac{i+j}{w}$,
where $j$ labels the chains stacked in the $y$ direction. The other parameters are $t_y=t_x$,  $L_{x}=10000$.}
\label{fig5} 
\end{figure}

In Fig.~\ref{fig3} we show the topological phase diagrams of the
SSG model at $L_{y}=6,7$ as functions of $m/t_{x}$ and $t_{d}/t_{x}$,
setting $t_{y}=t_{x}$. Surprisingly we find more phases than predicted
by the $\nu$ invariant. The gap closes not only for $m=0$  at $k=\pi$
when $L_{y}$ is odd but also for any $L_{y}$ along lines $m=m_{n}$ at $k=k_n$, where 
\begin{equation}
m_{n}=\frac{t_{d}t_{y}}{t_{x}}\varepsilon_{n}^{2},\ k_{n}=\pm2\arccos\left[\frac{t_{y}}{2t_{x}}\varepsilon_{n}\right], \ \varepsilon_{n}=2\cos\tfrac{n\pi}{L_{y}+1}\label{eq:linen}
\end{equation}
 and $n=1,2,\dots,\left\lfloor L_{y}/2\right\rfloor $
provided that 
\begin{equation}
\left|\frac{t_{y}}{2t_{x}}\varepsilon_{n}\right|\leq1. \label{hiddensymmetrycondition}
\end{equation}
This means that in the limit of very wide ribbon ($L_{y}\to\infty$)
the phase diagram consists of a quasi-continuous set of lines $t_{d}=\gamma_{n}m$
with slopes $\gamma_{n}$ ranging between $\tfrac{t_{x}}{4t_{y}}$
and $\infty$. The natural question to ask now is what is the origin
of these gap-closing lines? The answer are the hidden symmetries,
that can be found at the magical $k_{n}$ points, yielding to new 
$\mathbb{Z}_{2}$ invariants $\nu_{n}$. 

To see the hidden symmetries we rotate $\sigma_{\alpha}$
matrices by angle $\tfrac{k}{2}$ around the $z$-axis and use the eigenbasis of $\tau_{x}$ to transform the operators $\tau_{\alpha}$ in a block-diagonal form, where the blocks are given by (see Appendix
\ref{sec:Hidden-symmetries}) 
\[
\tau_{x,n}=\varepsilon_{n}\sigma'_{x},\quad\tau_{y,n}=\varepsilon_{n}\sigma'_{y},\quad\tau_{z,n}=\sigma'_{z}.
\]
and $\sigma'_{\alpha}$ is a new set of Pauli matrices. For odd $L_{y}$ the
blocks $n=0$ is given by $\tau_{x,0}=\tau_{y,0}=0$
and $\tau_{z,0}=1$. After this transformation the Hamiltonian (\ref{eq:hamk})
also has a block-diagonal form 
\begin{eqnarray}
{\cal H}'_{k,n} & = & -m\sigma_{z}\sigma'_{z}+2t_{x}\cos\tfrac{k}{2}\sigma_{x}\nonumber \\
 & + & 2t_{d}\varepsilon_{n}\cos\tfrac{k}{2}\sigma_{y}\sigma'_{y}+t_{y}\varepsilon_{n}\sigma'_{x}.\label{eq:hamk3}
\end{eqnarray}
Now we notice that ${\cal H}'_{k,n}$ is invariant under interchange
of $\sigma$ and $\sigma'$ operators if $2t_{x}\cos\tfrac{k}{2}=t_{y}\varepsilon_{n}$
which provides the condition for gap closing points in the $k$-space [Eq.~(\ref{eq:linen})].
The spin-interchange $X_{12}\vec{\sigma}X_{12}=\vec{\sigma}'$ and vice-versa is realized by  operator  $X_{12}=\tfrac{1}{2}(1+\vec{\sigma}\cdot\vec{\sigma}')$ 
 \cite{Brz14}.
The spectrum of $X_{12}$
consists of single $-1$ (singlet state) and three $+1$ (triplet states) eigenvalues. Thus ${\cal H}'_{k,n}$
in the eigenbasis of $X_{12}$ becomes block-diagonal with one block
being $1\times1$ and the other being $3\times3$. Therefore we can define a topological
$\mathbb{Z}_{2}$ invariant based on the sign of the matrix element of the $1\times1$ block. It changes at the gap closing
lines defined by Eq. (\ref{eq:linen}) so that it takes the form 
\begin{equation}
\nu_{n}=\tfrac{1}{2}\{1+{\rm sign}\:[m-m_{n}]\}.
\end{equation}
We conclude that the full topological description  the SSG model is given by a vector $\{\nu,\nu_{1},\nu_{2},\dots,\nu_{\left\lfloor L_{y}/2\right\rfloor }\}$
because changes of these invariants coincide with all the gap closing
lines in the phase diagram. Each hidden symmetry $n=1,2,\dots,\left\lfloor L_{y}/2\right\rfloor$ exists only if condition (\ref{hiddensymmetrycondition}) is satisfied. Therefore, it is possible to connect phases with different $\nu_n$ without closing the bulk gap by choosing a path in the parameter space which goes outside the region where the hidden symmetry exists [see Fig.~\ref{fig3}(c)], distinguishing the hidden-symmetric topological phases from the ones protected by structural symmetries.
Note that in the two-dimensional limit where $t_x=t_y$ and the system is periodic in both directions the hidden symmetries become a mirror symmetry with respect to the $\hat{x}+\hat{y}$ line, but the hidden symmetries can exist even if the mirror symmetry is absent (see Appendix \ref{sec:Hidden-symmetries2D}).

After establishing topological properties of the SSG ribbon we now turn our attention to the bulk-boundary correspondence. In Ref.~\cite{Sato15} it was argued that a NS chiral $\mathbb{Z}_{2}$ invariant does not generically support
end states in one dimension because the boundary necessarily breaks
the $S_{k}$ symmetry. This however does not exclude a special type
of smooth DWs from having zero energy bound states. To obtain analytical insights we can develop a continuum model for odd $L_{y}$ by
expanding ${\cal H}{}_{k}$ around gap closing point at $k=\pi$ and
get 
\begin{equation}
{\cal H}_{{\rm eff}}=v\delta k\sigma_{x}-m\sigma_{y} \ (v>0). 
\label{dirac}
\end{equation}
Now we create a DW of width $w$ in the real space 
\begin{equation}
 m(x)=\tfrac{m_{2}+m_{1}}{2}+\tfrac{m_{2}-m_{1}}{2}\tanh\tfrac{x}{w}
\end{equation}
between regions with positive mass $m_1=m_0$ and
negative mass $m_2=-m_0$  ($m_0>0$), separating phases with $\nu=1$ and $\nu=0$ [Figs.~\ref{fig2}(b) and \ref{fig4}(a)], and we find that a zero-energy eigenstate of ${\cal H}_{{\rm eff}}$ exists in a form
\begin{equation}
\psi(x)=(\begin{array}{cc}
0 & (\cosh\tfrac{x}{w})^{-m_0w/v} \end{array})^{T}/{\cal N},
\end{equation}
where ${\cal N}$ is a normalization factor.
This however does not take into account the fact that the chiral symmetry
of ${\cal H}_{{\rm eff}}$ becomes NS if one goes beyond linear order
in $\delta k$. Therefore, we implemented numerically such a DW, 
for which $|m|$ is constant within a unit cell as follows from Eq.~(\ref{dirac}),
and calculated the energy of the DW state as a 
function of $w$ [Figs.~\ref{fig2}(c)(d) and \ref{fig4}(a)]. This way we find that the energy of the topological DW state approaches zero as $1/w$ whereas the energies of the other bound states  scale as $1/\sqrt{w}$.
This  means that the topological DW state can be distinguished from other bound states based on the scaling behavior because its energy approaches zero faster than the energies of the other states. 
We emphasize that the scaling of the energy to zero as $1/w$ or faster is  a robust property of these topological DW states.  In the continuum model this state would have zero-energy  and the lattice effects can give maximally a correction proportional to $1/w$.  The robusness against perturbations preserving the NS chiral symmetry is demonstrated in Appendix \ref{sec:Robust}.
On the other hand, in the Appendix \ref{sec:Local} we show that the DW state is exponentially localized around the center of the DW, i.e. such position $x=x_0$ that $m(x_0)=0$, and the localization length decreases with increasing $w$, as one could expect.
The $1/\sqrt{w}$ scaling of the bulk gap can be
understood as well. By inserting an expansion $m(x)\simeq m_0 \tfrac{x}{w}$ around $x=0$   to Hamiltonian  ${\cal H}_{{\rm eff}}$ and eliminating $\psi_{1}$ we obtain a Harmonic oscillator equation for $\psi_2$
\begin{equation}
-v^2 \psi_{2}''(x)+\frac{m_0^2}{w^2}x^{2}\psi_{2}(x)=\left(E^2+\tfrac{m_0 v}{w}\right)\psi_{2}(x).
\end{equation}
The energies of this problem are given by
\begin{equation}
E_n=\sqrt{\frac{2n m_0 v}{w}}, \ n=0,1,2,...
\end{equation}
The $n=0$ solution gives the topological DW state and the energies of the other states scale as $1/\sqrt{w}$. 
We have just shown that the in-gap state at the DW between two topologically distinct
domains follows from the continuum-limit model and adiabatic evolution of the states when DW is sufficiently smooth.
This is quite a different case to the one discussed in \cite{Top17} where a surface state also appears in a presence of a NS 
symmetry but it also requires a surface potential.

We find even more striking bulk-boundary correspondence for phases
described by $\nu_{n}$ invariants. In Figs.~\ref{fig2}(d) and \ref{fig4}(c),(d)
we show the scaling of energies of the topological DW states and the non-topological states as a function of $w$ when the masses 
 $m_{1}$ and $m_{2}$ are chosen so that 
the DW separates two different $\nu_{n}$ phases. The energies of the non-topological states behave in the same way as before, scaling
as $1/\sqrt{w}$, but the energies of the hidden-symmetry protected topological DW states scale as $e^{-w/\xi}$. By expanding Hamiltonian around the gap closing points $\pm k_n$  we get two
similar bound state solutions as for $k=\pi$. However, these solutions do not give zero-energy states even in the continuum model  because the gap closes at two different momenta so that these bound states hybridize leading to non-zero energy. Nevertheless, in Appendix \ref{sec:Overlap-of-two}
we show using the properties of the Schwartz functions that in the continuum model their
overlap vanishes exponentially fast with $w$ which allows the possibility of the exponential scaling. Nevertheless, the lattice effects could lead to $1/w$ corrections also in this case. Therefore, we have studied more carefully how the scaling of the energies depends on the details of the DW. We find that  for a specific type of DW   it is possible to obtain an exponential decay of energy $e^{-w/\xi}$ also in the case of the DWs separating different $\nu$ phases,  and conversely it is possible to obtain $1/w$ scaling for a DW separating different $\nu_n$ phases (see Fig.~\ref{fig5}). 
In this way, by manipulating the details of the DWs, we are able to see that the behavior of both types of DW states are equivalent.
This leads to a robust conclusion that the energies of the DW states scale as $1/w$ or faster. While the $1/w$ scaling is expected to be generic based on the analytical arguments given above, the faster exponential scaling, which we numerically find in some specific circumstances, means that the $1/w$ corrections are not always present. The analytical model-independent theoretical understanding of the conditions for the exponential scaling is an interesting direction for future research. By setting $m_{1}$ and
$m_{2}$ is such a way that all the gap closing lines are crossed
on the way from $m_{1}$ to $m_{2}$ we can always obtain extensive
number of DW states $L_{y}$  both for even and odd $L_{y}$.

An interesting proterty of the SSG model is that when the width of the DW $w$ increases a single state separates
from the bulk spectrum and tends to zero from above or below [Fig. \ref{fig2}(c)]. This asymmetric spectral flow needs to be taken into account when calculating the charges for solitons and antisolitons  (see Appendix \ref{sec:Charge-of-a}). For $L_y=1$ we obtain $q_0=\frac{\pm1-\delta}{2}, \frac{\pm1+\delta}{2}$ for solitons and antisolitons depending on whether the zero-energy  state is occupied or empty. Here $\delta=\frac{\zeta_2-\zeta_1}{2}$,
\begin{equation}
\zeta_i=\frac{2}{\pi}\frac{m_i}{\sqrt{m_i^{2}+4t_{x}^{2}}}K\left(\frac{4t_{x}^{2}}{m_i^{2}+4t_{x}^{2}}\right) \nonumber \label{eq:delta}
\end{equation}
are the differences of the bulk filling factors of the two sublattices in the region with mass $m_{1}>0$ and $m_2<0$
and $K(x)$ is the complete elliptic integral of the first kind. The DWs between different hidden-symmetric topological phases carry charges $q_n=0, \pm 1$, so that for a general DW the charge is the sum of   $q_0$  and the charges $q_n$ contributed by the transverse modes supporting transitions between different hidden-symmetric topological phases.

To summarize, we have analytically described the topological properties of the SSG model and propose it as a paradigmatic model for NS chiral-symmetric topological phases. We have shown that a smooth
DW supports zero energy state(s) if the DW separates regions
with different NS chiral invariant $\nu$ or different hidden-symmetry
invariants $\nu_n$.  In addition to engineered artificial lattices \cite{Kitagawa12, Atala13, Wang13, Lohse16, Nakajima16, Leder16, Meier16, Yan19, Drost17, Huda18, Fischer18, Fasel18}  our findings are also relevant  in the context of low-dimensional binary compounds
supporting surface atomic steps. In these systems the surface steps lead to one-dimensional topological modes and the system obeys NS chiral symmetry, so that DWs between topologically distinct phases can support  DW states \cite{Brz18}, providing a possible explanation for the zero-bias conductance peak observed in the recent experiment \cite{Mazur}.

\begin{acknowledgments}
The work is supported by the Foundation for Polish Science through
the IRA Programme co-financed by EU within SG OP Programme.
W.B. also acknowledges support by Narodowe Centrum Nauki 
(NCN, National Science Centre, Poland) Project No. 2016/23/B/ST3/00839.
\end{acknowledgments}

\appendix

\section{Hamiltonian and its symmetries\label{sec:Symmetries}}

The SSG Hamiltonian on a square lattice has a form
\begin{eqnarray}
{\cal H} & = &t_{x}\sum_{\vec{j}} (c_{\vec{j}}^{\dagger}c_{\vec{j}+\hat{x}}+h.c)+  t_{y}\sum_{\vec{j}} (c_{\vec{j}}^{\dagger}c_{\vec{j}+\hat{y}}+h.c.) \nonumber\\&&+t_{d}\sum_{s=\pm1}\sum_{\vec{j}}(-1)^{j_x+j_y+1}(c_{\vec{j}}^{\dagger}c_{\vec{j}+\hat{x}+s\hat{y}}+h.c.)\nonumber\\ &&+ m\sum_{\vec{j}}(-1)^{j_x+j_y+1}c_{\vec{j}}^{\dagger}c_{\vec{j}}
\end{eqnarray}
where $m$ is the mass term, $t_{x}$
and $t_{y}$ are hopping amplitudes along $x$
and $y$ directions and $t_{d}$ is the diagonal hopping amplitude.
The $k$-space form is given by
\begin{eqnarray}
{\cal H}_{k} & = & -m\sigma_{z}\tau_{z}+2t_{x}\cos\tfrac{k}{2}(\cos\tfrac{k}{2}\sigma_{x}-\sin\tfrac{k}{2}\sigma_{y})\nonumber \\
 & + & t_{y}\tau_{x}+2t_{d}\cos\tfrac{k}{2}(\sin\tfrac{k}{2}\sigma_{x}+\cos\tfrac{k}{2}\sigma_{y})\tau_{y},
\end{eqnarray}
where $\tau_{\alpha}$ operators are given by matrices
\begin{equation}
\tau_{x}=\left(\begin{array}{ccccc}
0 & 1 & 0 & 0 & \cdots\\
1 & 0 & 1 & 0 & \cdots\\
0 & 1 & 0 & 1\\
0 & 0 & 1 & 0\\
\vdots & \vdots &  &  & \ddots
\end{array}\right),\;\tau_{y}=\left(\begin{array}{ccccc}
0 & -i & 0 & 0 & \cdots\\
i & 0 & i & 0 & \cdots\\
0 & -i & 0 & -i\\
0 & 0 & i & 0\\
\vdots & \vdots &  &  & \ddots
\end{array}\right),
\end{equation}
and
\begin{equation}
\tau_{z}=\left(\begin{array}{ccccc}
1 & 0 & 0 & 0 & \cdots\\
0 & -1 & 0 & 0 & \cdots\\
0 & 0 & 1 & 0\\
0 & 0 & 0 & -1\\
\vdots & \vdots &  &  & \ddots
\end{array}\right),\;\tau_{a}=\left(\begin{array}{ccccc}
\cdots & 0 & 0 & 0 & 1\\
\cdots & 0 & 0 & 1 & 0\\
 & 0 & 1 & 0 & 0\\
 & 1 & 0 & 0 & 0\\
\iddots &  &  & \vdots & \vdots
\end{array}\right).
\end{equation}
Here we defined additional matrix $\tau_{a}$ which is needed to construct
some of the symmetry operators.

Depending on system width $L_{y}$ being even or odd the system have
different symmetry properties. However, some symmetries are common for
both cases. The one which is most relevant here is the non-symmorphic
(NS) chiral symmetry defined as $S_{k}=\sin\tfrac{k}{2}\sigma_{x}\tau_{z}+\cos\tfrac{k}{2}\sigma_{y}\tau_{z}$
which satisfies $S_{k}{\cal H}_{k}S_{k}^{-1}=-{\cal H}_{k}$.
The $k$-dependence in $S_{k}$ is intrinsic and follows from the
half lattice translation that is needed to go from one sublattice
to the other. We also have a time-reversal
symmetry for spinless particles ${\cal T}{\cal H}_{k}{\cal T}^{-1}={\cal H}_{-k}$, where  ${\cal T}={\cal K}$ is complex
conjugation. Finally,
for any $L_{y}$ we have a symmetry with respect to a mirror line
perpendicular to $x$ direction, passing through a lattice site, taking
a form of $M_{x}=\cos\tfrac{k}{2}-i\sin\tfrac{k}{2}\sigma_{z}$ and
acting as $M_{x}{\cal H}_{k}M_{x}^{-1}={\cal H}_{-k}$. Despite the
$k$-dependence this is a symmorphic symmetry. By shifting a mirror
line to cut a middle of a bond we can also get a chiral mirror symmetry
$\overline{M}_{x}=\sigma_{y}\tau_{z}$ yielding relation $\overline{M}_{x}{\cal H}_{k}\overline{M}_{x}^{-1}=-{\cal H}_{-k}$.

For odd $L_{y}$ we have another mirror symmetry with respect to line perpendicular to $y$ direction $M_{y}{\cal H}_{k}M_{y}^{-1}={\cal H}_{k}$, where
$M_{y}=\tau_{a}$.
For even $L_{y}$ mirror $M_{y}$ does not exist but we have a particle-hole
symmetry ${\cal C}=i{\cal K}\sigma_{z}\tau_{a}\tau_{z}$ and inversion
symmetry $I=\sigma_{x}\tau_{a}$, yielding relations ${\cal C}{\cal H}_{k}{\cal C}^{-1}=-{\cal H}_{-k}$
and $I{\cal H}_{k}I^{-1}={\cal H}_{-k}$.

It is important to notice that in the eigenbasis of  $S_{k}$ the mirror symmetry operator $M_{x}$ 
operator transforms to identity. This is possible because if we put eigenvectors of  $S_{k}$ in the columns
of unitary matrix ${\cal U}_k$ then  $M_{x}$ is transformed as $\tilde{M}_{x}={\cal U}_{-k}^{\dagger}M_{x}{\cal U}_k$.
Note that this is not similarity transformation so the spectrum of $M_{x}$ is not left invariant. The form of 
transformation is dictated by the mirror-symmetry relation with the Hamiltonian in the new basis. Denoting
$\tilde{{\cal H}}_{k}={\cal U}_{k}^{\dagger}M_{x}{\cal U}_k$ we get that $\tilde{M}_{x}\tilde{{\cal H}}_{k}\tilde{M}_{x}^{-1}=\tilde{{\cal H}}_{-k}$.

\section{Hidden symmetries\label{sec:Hidden-symmetries}}

To see the hidden symmetries we first transform ${\cal H}_{k}$ to ${\cal H}'_{k} =  R_{k/2}^{\dagger}{\cal H}_{k}R_{k/2}$ using
$R_{k/2}=\exp(i\tfrac{k}{4}\sigma_{z})$ to get
\begin{equation}
{\cal H}'_{k} = -m\sigma_{z}\tau_{z}+2t_{x}\cos\tfrac{k}{2}\sigma_{x}
 +  2t_{d}\cos\tfrac{k}{2}\sigma_{y}\tau_{y}+t_{y}\tau_{x}.  \label{eq:hamk2-1}
\end{equation}
The eigenfunctions of $\tau_{x}$ corresponding to eigenvalues $\varepsilon_{n}$ are $\psi_{n}(j)=\sqrt{\tfrac{2}{L_{y}+1}}\sin\tfrac{nj\pi}{L_{y}+1}$,
where $j$ labels sites in the $y$ direction and $n$ labels modes
($j,n=1,2,\dots,L_{y}$). We can use these transverse modes to construct a new basis 
 $\left|\phi_{2n-1}\right\rangle =\left(\left|\psi_{n}\right\rangle +\left|\psi_{L_{y}+1-n}\right\rangle \right)/\sqrt{2}$
and $\left|\phi_{2n}\right\rangle =\left(\left|\psi_{n}\right\rangle -\left|\psi_{L_{y}+1-n}\right\rangle \right)/\sqrt{2}$
for $n=1,2,\dots,\left\lfloor L_{y}/2\right\rfloor $ and if $L_{y}$
is odd $\left|\phi_{0}\right\rangle =\left|\psi_{\left\lceil L_{y}/2\right\rceil }\right\rangle$. In this basis $\tau_{x}$, $\tau_{y}$ and $\tau_{z}$ have block-diagonal
forms with $\left\lfloor L_{y}/2\right\rfloor $ diagonal blocks given
by
\[
\tau_{x,n}=\varepsilon_{n}\sigma'_{x},\quad\tau_{y,n}=\varepsilon_{n}\sigma'_{y},\quad\tau_{z,n}=\sigma'_{z},
\]
where $\sigma'_{\alpha}$ is a new set of Pauli matrices, and for odd
$L_{y}$ the block $n=0$ is given
by $\tau_{x,0}=\tau_{y,0}=0$ and $\tau_{z,0}=1$. Thus
the Hamiltonian  has a block-diagonal form
\begin{eqnarray}
{\cal H}'_{k,n} & = & -m\sigma_{z}\sigma'_{z}+2t_{x}\cos\tfrac{k}{2}\sigma_{x}\nonumber \\
 & + & 2t_{d}\varepsilon_{n}\cos\tfrac{k}{2}\sigma_{y}\sigma'_{y}+t_{y}\varepsilon_{n}\sigma'_{x} \label{eq:hamk3-1}
\end{eqnarray}
supporting the hidden symmetries discussed in the main text.

\section{Hidden symmetries in a two-dimensional limit \label{sec:Hidden-symmetries2D}}

In case when the system is periodic both in $x$ and $y$ direction it can be described by a 
$k$-space Hamiltonian written in a form of
\begin{eqnarray}
{\cal H}_{\vec{k}} & = & -m\sigma_{z}\tau_{z}+2t_{x}\cos\tfrac{k_x}{2}(\cos\tfrac{k_x}{2}\sigma_{x}-\sin\tfrac{k_x}{2}\sigma_{y})\nonumber \\
 & + & 2t_{y}\cos\tfrac{k_y}{2}(\cos\tfrac{k_y}{2}\tau_{x}-\sin\tfrac{k_y}{2}\tau_{y})\nonumber \\
 & + & 4t_{d}\cos\tfrac{k_x}{2}\cos\tfrac{k_y}{2}(\sin\tfrac{k_x}{2}\sigma_{x}+\cos\tfrac{k_x}{2}\sigma_{y})\nonumber \\
 & &(\sin\tfrac{k_y}{2}\tau_{x}+\cos\tfrac{k_y}{2}\tau_{y}),\label{eq:hamk}
\end{eqnarray}
where $\sigma_{\alpha}$ and $\tau_{\alpha}$ are the Pauli matrices describing the unit cell 
along $x$ and $y$ directions. Similarly as before, to see the hidden symmetries we transform
${\cal H}_{\vec{k}}$ to ${\cal H}'_{\vec{k}} =  R_{\vec{k}}^{\dagger}{\cal H}_{\vec{k}}R_{\vec{k}}$ using
$R_{\vec{k}}=\exp(\tfrac{i}{4}[k_x\sigma_{z}+k_y\tau_{z}])$ to get
\begin{eqnarray}
{\cal H}'_{\vec{k}} &=& -m\sigma_{z}\tau_{z}+2t_{x}\cos\tfrac{k_x}{2}\sigma_{x}+2t_{y}\cos\tfrac{k_y}{2}\tau_{x}\nonumber \\
 &+& 4t_{d}\cos\tfrac{k_x}{2}\cos\tfrac{k_y}{2}\sigma_{y}\tau_{y}.  
\label{eq:hamk4-1}
\end{eqnarray}
Obviously ${\cal H}'_{\vec{k}}$ (and already  ${\cal H}_{\vec{k}}$) is invariant under interchange
of $\sigma$ and $\tau$ operators if 
\begin{equation}
t_{x}\cos\tfrac{k_x}{2}=t_{y}\cos\tfrac{k_y}{2}.
\label{cond}
\end{equation}
The spin-interchange $X_{12}\vec{\sigma}X_{12}=\vec{\sigma}'$ and vice-versa is realized 
by  operator  $X_{12}=\tfrac{1}{2}(1+\vec{\sigma}\cdot\vec{\tau})$. Coming back to the original
basis of Eq.(\ref{eq:hamk}) we find that $X_{12}$ operator gives rise to two different symmetry operations,
\begin{eqnarray}
M_{xy}&\equiv& R_{k_y,k_x}X_{12}R_{k_x,k_y}^{\dagger}=\tfrac{1}{2}(1+\vec{\sigma}\cdot\vec{\tau}), \\
{\cal X}_{\vec{k}}&\equiv& R_{k_x,ky}X_{12}R_{k_x,k_y}^{\dagger}=\tfrac{1}{2}(1+\cos\tfrac{k_x-k_y}{2}[\sigma_x\tau_x+\sigma_y\tau_y]  \nonumber\\
&+&\sin\tfrac{k_x-k_y}{2}[\sigma_x\tau_y-\sigma_y\tau_x]+\sigma_z\tau_z).
\end{eqnarray}
The first one is the mirror symmetry with respect to the $\hat{x}+\hat{y}$ line. One can verify that 
if $t_{x}=t_{y}$ then 
\begin{equation}
M_{xy}{\cal H}_{k_x,k_y}M_{xy}^{-1}={\cal H}_{k_y,k_x}.
\end{equation}
Therefore $M_{xy}$ commutes with ${\cal H}_{\vec{k}}$ for $k_x=k_y$. Note that this also holds 
for a finite periodic system where $L_x\not=L_y$,  when the discretized quasimomentum satisfies $k_x=k_y$.
On the other hand, the hidden symmetry operator ${\cal X}_{\vec{k}}$  satisfies
\begin{equation}
{\cal X}_{k_x,f(k_x)}{\cal H}_{k_x,f(k_x)}{\cal X}_{k_x,f(k_x)}^{-1}={\cal H}_{k_x,f(k_x)},
\end{equation}
as long as condition (\ref{cond}) holds which means that 
\begin{equation}
k_y=f(k_x)=2\arccos\left(\frac{t_x}{t_y}\cos\frac{k_x}{2}\right).
\end{equation}
Note that this does not require that $t_{x}=t_{y}$. Hence, the hidden symmetry is a different operator than
the mirror symmetry $M_{xy}$ eventhough it originates from the same $X_{12}$ operator in the
$R_{\vec{k}}$--transformed basis. It can be regarded as an interchange of longitudinal and transverse modes
in the system. Note that condition (\ref{cond}) trivially generalizes to the condition found for a multimode wire.

\section{Robustness of the DW states\label{sec:Robust}}

\begin{figure}[t]
\begin{centering}
\includegraphics[width=1\columnwidth]{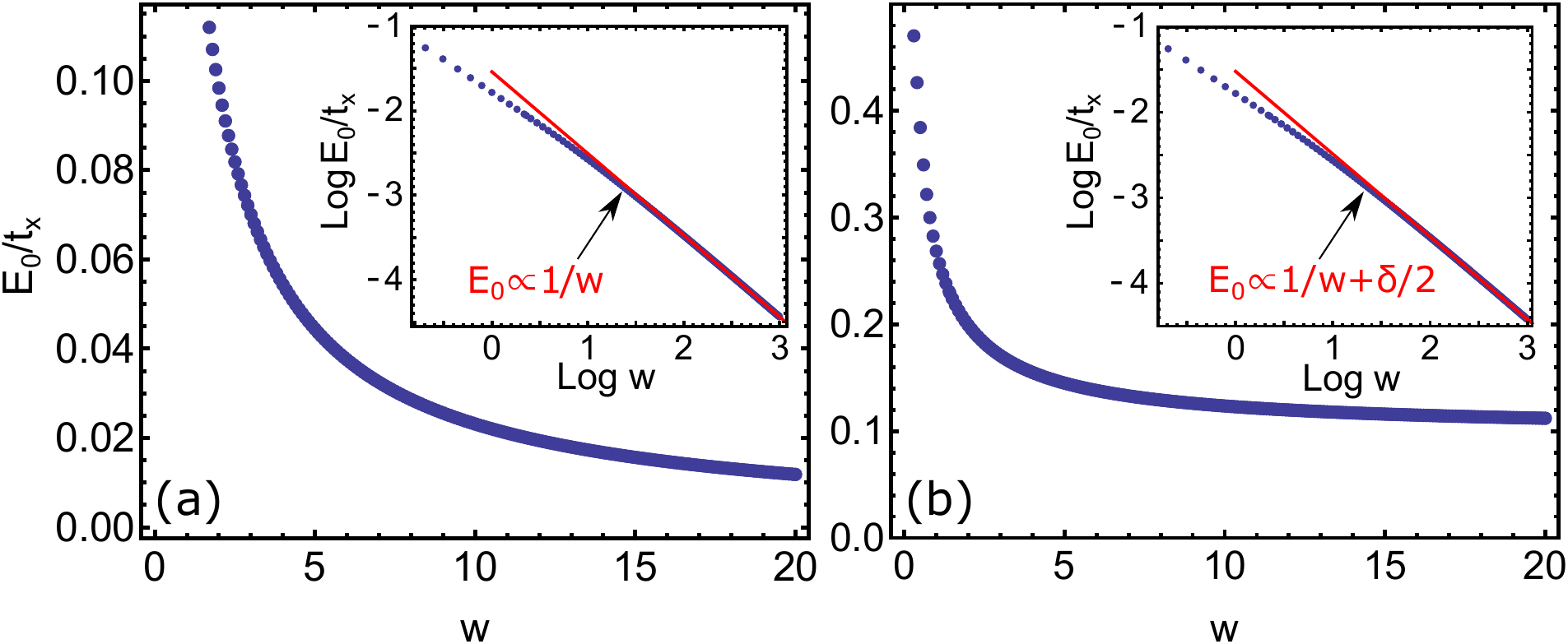} \vspace{-0.6cm}
\par\end{centering}
\caption{Scaling of the energy of the topological DW state $E_{0}$ of a single-mode SSG model as a function of the DW width $w$ in linear and logarithmic scales (insets). 
(a) We have added a generic perturbation that preserves NS chiral symmetry. (b) We have added a perturbation that breaks NS symmetry with $\delta=0.1$. 
The parameters are $m_{1,2}=\pm t_x$ and $L_x=10000$. \label{fig6} }
\end{figure}
To check whether the DW states protected by the NS chiral symmetry are indeed topologically protected we 
consider a single-mode SSG wire. An arbitrary perturbation that preserve the NS chiral symmetry ${\cal S}_k$ 
has a form of
\begin{equation}
{\cal V}_{k} =  a_k \sigma_z + b_k(\cos\tfrac{k}{2}\sigma_x-\sin\tfrac{k}{2}\sigma_y),
\label{Vk}
\end{equation}
where $k$-dependent coefficients should have such form that ${\cal V}_{k}$ is $2\pi$-periodic in $k$. Hence we have
\begin{eqnarray}
a_k &=& p_0 + \sum_{i=1}^r(p_{2i-1}\sin(ik)+p_{2i}\cos(ik)),\nonumber\\
b_k &=& \sum_{i=1}^r(q_{2i-1}\sin\tfrac{(2i-1)k}{2}+q_{2i}\cos\tfrac{(2i-1)k}{2}),
\end{eqnarray}
where $r$ is  the range of hoppings involved in ${\cal V}_{k}$. 

In Fig.~\ref{fig6} we show the scaling of the energy of the DW state between two regions of 
opposite masses for a single-mode wire in presence of perturbation ${\cal V}_{k}$. We choose
$r=10$ and we pick random $p_n$ and $q_n$ coefficients from the uniform distribution over
the interval $[-0.1t_x,0.1t_x]$. The masses are chosen as $m_{1,2}=\pm t_x$  so that such perturbation
does not close the bulk gap. As we can see in Fig.~\ref{fig6}(a) the energy $E_0$ of the DW state still scales
to zero as $1/w$. On the other hand, if we add a perturbation that breaks the NS chiral symmetry, for
instance ${\cal V}_{k} = \delta\sigma_x$, the DW state should converge to a non-zero energy. In
Fig.~\ref{fig6}(b) we see that this is indeed the case for our single-mode wire $E_0$ scales like $1/w$ to
$\delta$.

\section{Localization of the DW states\label{sec:Local}}

\begin{figure}[t]
\begin{centering}
\includegraphics[width=1\columnwidth]{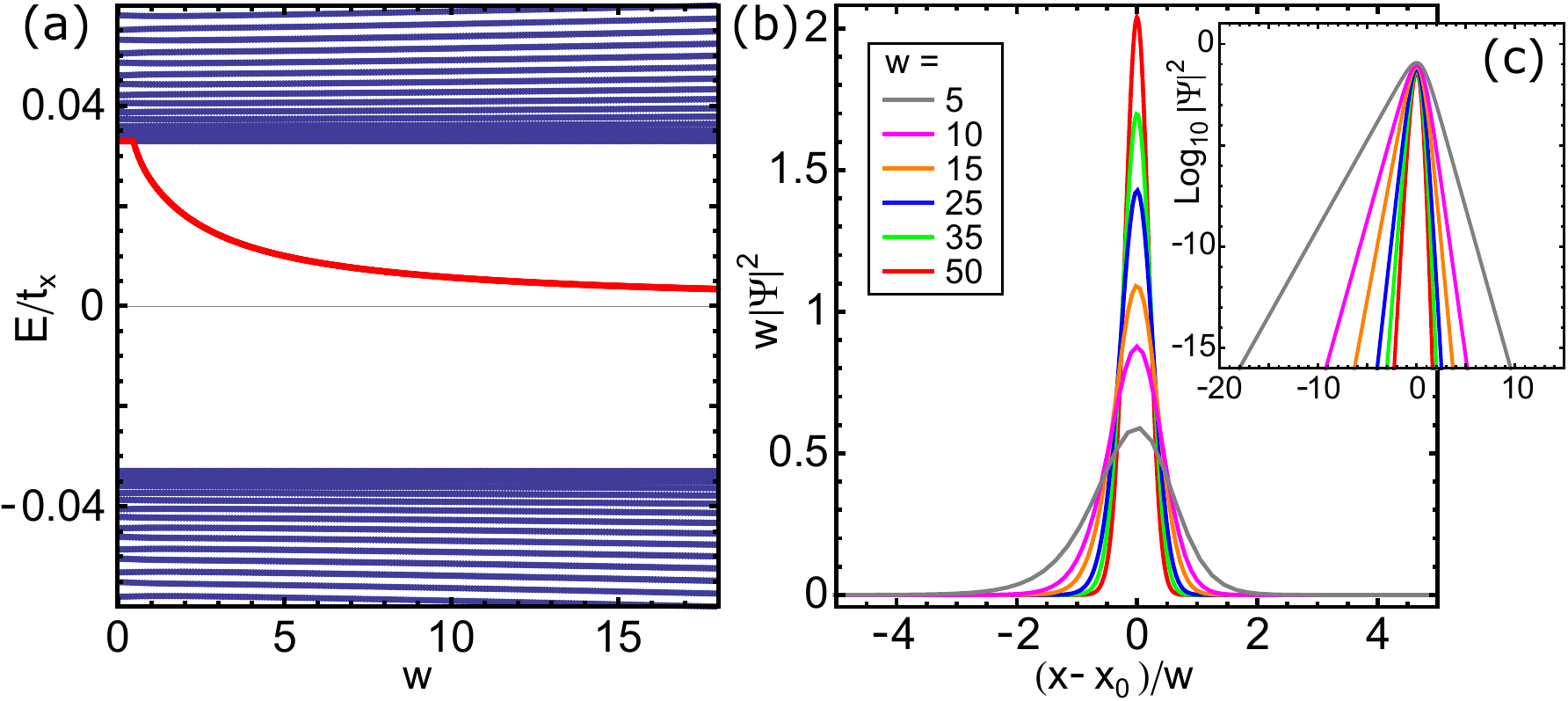} \vspace{-0.6cm}
\par\end{centering}
\caption{
(a) Spectral flow of the DW states for multimode SSG chain as function of the DW width $w$ for $L_y=7$, $L_x=1000$ and $t_y=t_x$ . DW separates regions
with masses $m_1=0.2 t_x$ and $m_2=-0.4 t_x$ and $t_d=0.4 t_x$. (b) Evolution of local density of states for the DW state for increasing $w$, $x$ is the position along
the chain (lines are continously interpolated between the chain's sites as a guide for the eye) and $x_0$ is the center of the DW defined by $m(x_0)=0$. Horizona axis is
renormalized by $w$ and vertical by $1/w$. (c) The linear-log version of plot (b).
 \label{fig7}}
\end{figure}

In Fig. \ref{fig7} we show the spectral flow and the local density of states for a DW state in a representative multimode chain with $L_y=7$ for increasing
width $w$ of a domain wall between two domains with different NS $\mathbb{Z}_2$ invariant. We note that the DW state is exponentially localized around
the center of the domain wall at $x=x_0$ (defined by $m(x_0)=0$) and the localization length decreases with increasing $w$ .

\section{Fermi velocity at Dirac points $k=\pm k_{n}$\label{sec:Fermi-veolocity-at}}

We notice that the Hamiltonian ${\cal H}'_{k,n}$ given by Eq. (\ref{eq:hamk3})
commutes with ${\cal P}=\sigma_{x}\sigma'_{x}$. The bands that cross
at $k=\pm k_{n}$ can be found in the block ${\cal P}=-1$ of the
Hamiltonian that takes the form,

\begin{eqnarray}
{\cal H}'_{k,n,-} & = & -4\sin\frac{k+k_{n}}{4}\sin\frac{k-k_{n}}{4}\left(t_{x}\sigma_{x}+t_{d}\varepsilon_{n}\sigma_{z}\right)\nonumber \\
 & - & \left(m-m_{n}\right)\sigma_{z}.
\end{eqnarray}
The remaining ${\cal P}=+1$ block is related by a shift of $2\pi$
in the $k$-space and unitary transformation, namely ${\cal H}'_{k,n,+}=\sigma_{z}{\cal H}'_{k+2\pi,n,-}\sigma_{z}$.
Consequently, the Fermi velocity at the Dirac points is given by
\begin{equation}
v_{n}=\pm t_{x}\sqrt{\left(1-\frac{t_{y}^{2}}{4t_{x}^{2}}\varepsilon_{n}^{2}\right)\left(1+\frac{t_{d}^{2}}{t_{x}^{2}}\varepsilon_{n}^{2}\right)}.
\end{equation}

\section{Overlap of two domain-wall functions\label{sec:Overlap-of-two}}

We assume that the functional form of $m(x)$ is such that it only depends on parameter $x/w$. In analogy to the domain wall solution for a gap closing point at
$k=\pi$ we get two solutions for $k=\pm k_{n}$ gap closings in a
form (we ignore the spinor structure which would change as a function momentum and the normalization factor which is not important for the statements below)
\begin{equation}
\psi^{\pm}(x)=\exp\left[\pm ik_{n}x-v_{n}^{-1}\int_0^x m\left(\frac{x'}{w}\right)dx'\right].
\end{equation}
Their overlap is
\begin{equation}
\langle\psi^{-}|\psi^{+}\rangle=\int_{-\infty}^{+\infty}\exp\left[2ik_{n}x-2v_{n}^{-1}\int_0^x m\left(\frac{x'}{w}\right)dx'\right]dx
\end{equation}
and substituting $x=yw$ we get
\begin{equation}
\langle\psi^{-}|\psi^{+}\rangle=w\int_{-\infty}^{+\infty}\exp\left[2ik_{n}wy-2wv_{n}^{-1}M(y)\right]dy,
\end{equation}
where $M(y)=\int_0^y m(y')dy'$. Assuming that $M(y)>0$,  $m(\infty)>0$ and $m(-\infty)<0$ as we expect from a domain wall, we
notice that   $f(y)=\exp\left[-2v_{n}^{-1} w M(y)\right]$ is a Schwartz function. This property does not depend on the details of the model (even if one goes beyond the linear order expansion in momentum) or the shape of the DW as long as the solutions $\psi_\pm$ are smooth functions decaying exponentially (or faster) far away from the DW.
The Fourier transform of a Schwartz function is also a Schwartz function and therefore the overlap $\langle\psi^{-}|\psi^{+}\rangle$
must vanish quicker than any power of $1/(k_{n}w)$. 

To see the exponential dependence explicitly we can assume for example $m(x/w)=m_0 x/w$. Then we obtain 
\begin{eqnarray}
\langle\psi^{-}|\psi^{+}\rangle&=&w\int_{-\infty}^{+\infty}\exp\left[2ik_{n}wy-wm_0v_{n}^{-1}y^2\right]dy \nonumber\\
&=&w \sqrt{\frac{\pi v_n}{wm_0}} \exp\left[-w k_{n}^2 v_n/m_0\right].
\end{eqnarray}

\section{Charge of a domain wall\label{sec:Charge-of-a}}

To understand the charge of the DW we start by calculating the charge appearing at the end of single-mode SSG chain with open boundary conditions
\begin{equation}
{\cal H}=\left(\begin{array}{ccccc}
-m & t & 0 & 0 & \cdots\\
t & m & t & 0 & \cdots\\
0 & t & -m & t & \cdots\\
0 & 0 & t & m\\
\vdots & \vdots & \vdots &  & \ddots
\end{array}\right).\label{eq:ham-open}
\end{equation}
Assuming even number of sites $L=2N$ the Hamiltonian can be Block diagonalized into 2x2 blocks similarly as in Appendix \ref{sec:Hidden-symmetries}, and the eigenstates can be found by diagonalizing these blocks. This way we find that the eigenenergies are ($n=1,2,....,N$)
\begin{equation}
E_{n \pm} = \pm E_n, \ E_n=\sqrt{m^2+4 t^2 \cos^2\bigg(\frac{n \pi}{2N+1} \bigg)}.
\end{equation}
The eigenstate corresponding to $E_{n-}$ at lattice site $j$ is 
\begin{eqnarray}
\psi_{n, -}(j)&=&\frac{\sin(\frac{n j \pi}{2N+1})}{\sqrt{2(2N+1)}}\bigg[[1-(-1)^j] \sqrt{1+\frac{m}{E_n}} \nonumber\\
&&-[1+(-1)^j] \sqrt{1-\frac{m}{E_n}} \ \bigg]. 
\end{eqnarray}
The total end charge $q_{end}$ at lattice sites $j=1,...,2 \xi$ (relative to the corresponding bulk charge $\xi$) is
\begin{eqnarray}
q_{end}&=&\frac{2}{2N+1}\sum_{n=1}^N \sum_{j=1}^{2\xi} \sin^2(\frac{n j \pi}{2N+1}) [1-(-1)^j \frac{m}{E_n}]-\xi \nonumber \\ 
&=& \frac{1}{2N+1}\sum_{n=1}^N \sum_{j=1}^{2\xi} (-1)^j \cos\bigg(\frac{n j 2\pi}{2N+1}\bigg)  \frac{m}{E_n}  \nonumber\\
&=& -\frac{\zeta}{4}+\frac{1}{2}  \frac{1}{2N+1}\sum_{n=1}^N  \frac{\cos\big[(2\xi+\frac{1}{2}) \frac{n  2\pi}{2N+1} \big]}{\cos\big[\frac{1}{2}\frac{n  2\pi}{2N+1} \big]}  \frac{m}{E_n} \nonumber\\
&=&\frac{{\rm sign}(m)-\zeta}{4},
\end{eqnarray}
where
\begin{eqnarray}
\zeta(m) &=&  \frac{2}{2N+1}\sum_{n=1}^N \frac{m}{E_n} =  \frac{2}{2\pi} \int_0^\pi dk  \frac{m}{\sqrt{m^2+4 t^2 \cos^2(k/2)}}\nonumber \\
&=&\frac{2}{\pi}\frac{m}{\sqrt{m^2+4t^2}} K\bigg(\frac{4t^2}{m^2+4t^2}\bigg)
\end{eqnarray}
is the difference of the bulk filling factors of the two sublattices and
\begin{equation}
K(x)= \int_0^{\pi/2} d\theta  \frac{1}{\sqrt{1-x \sin^2(\theta)}}
\end{equation}
is the complete elliptic integral of the first kind. Here we have used 
\begin{equation}
\frac{1}{4\pi} \int_0^\pi dk  \frac{m \cos\big[(2\xi+\frac{1}{2}) k \big]}{\sqrt{m^2+4 t^2 \cos^2(\frac{k}{2})}\cos\big[\frac{1}{2}k \big]}  =\frac{{\rm sign}(m)}{4},
\end{equation}
which is valid up to corrections which decay exponentially with increasing $\xi$. At the other end of the chain there is a charge $-q_{end}$.

Let's now consider a sharp DW between regions with mass $m_i$ and $m_j$. If we first assume that we turn off the hopping connecting these two regions, we find that the region with mass $m_i$ gives rise to the charge $-q_{end, i}$ and the region with mass $m_j$ gives rise to a charge $q_{end, j}$ at the DW so that the total charge is
\begin{equation}
q_{DW}=\frac{{\rm sign}(m_j)-{\rm sign}(m_i)}{4}-\frac{\delta_{ji}}{2},
\end{equation}
where we have defined
\begin{equation}
\delta_{ji}=\frac{\zeta(m_j)-\zeta(m_i)}{2}.
\end{equation}
This charge is exponentially localized at the DW and therefore when the hopping connecting the regions is turned on it causes only a local perturbation in the Hamiltonian (slightly redistributing the charge density locally) but does not influence the total charge localized at the DW.
If both $m_i$ and $m_j$ have the same sign the regions are in the same topological phase and the $q_{DW}=-\delta_{ji}/2$. If the signs of the masses are different we have DW between two topologically distinct phases. We now consider two possible DWs: (i) Soliton where mass changes from $m_1>0$ to $m_2<0$ as a function of increasing $x$ and (ii) antisoliton where mass changes from $m_2<0$ to $m_1>0$ as a function of increasing $x$ (see Figs.~\ref{fig1} and \ref{fig2} in the main text.)  We denote $\delta=\delta_{21}$.

(i) In the case of soliton the sharp DW carries a charge  $q_{DW}=-(1+\delta)/2$. By increasing the width of the DW we find that the spectral flow is such that a state will approach zero energy from positive energies (see Fig.~\ref{fig2} in the main text). This means that if this zero-energy state is unoccupied the charge of the DW is $q_{DW}$ and if it is occupied the charge is $q_{DW}+1$. Thus we can summarize that the possible charges for soliton are 
\begin{equation}
q_{DW}=\frac{\pm 1 -\delta}{2}.
\end{equation}

(ii) For antisoliton the charge of the sharp DW is  $q_{DW}=(1+\delta)/2$. By increasing the width of the DW we find that the spectral flow is such that a state will approach zero energy from negative energies. This means that if this zero-energy state is unoccupied the charge of the DW is $q_{DW}-1$ and if it is occupied the charge is $q_{DW}$. Thus we can summarize that the possible charges for antisoliton are 
\begin{equation}
q_{DW}=\frac{\pm 1 +\delta}{2}.
\end{equation}

In the case of multimode SSG system we can separate the transverse modes $n$ as discussed in Appendix~\ref{sec:Hidden-symmetries}. We consider the cases (a) $L_y$ even and (b) $L_y$ odd separately.

(a) When $L_y$ is even the Hamiltonian can be decomposed into 4x4 blocks given by Eq.~(\ref{eq:hamk3-1}). Each of these blocks $n=1,...L_y/2$ supports symmorphic chiral symmetries
$C_{yz}=\sigma_y \sigma_z'$ and $C_{zy}=\sigma_z \sigma_y'$. Therefore using the argument given in Ref.~\cite{Hou07} we find that each transverse mode $n$ carries possible charges $q_n=-N_n/2,...,N_n/2$, where $N_n$ is the number of zero-energy states at the DW supported by transverse mode $n$ and the value of charge $q_n$ is determined by the number of occupied zero-energy states. In the case of smooth DWs each transverse mode supports $N_n=2$ zero-energy states if $\nu_n$ is different on the two sides of the DW  so that $q_n=-1,0,1$ (with two possible states corresponding to $q_n=0$). If $\nu_n$ is the same on both sides then $N_n=0$ and $q_n=0$. The total charge of the DW is 
\begin{equation}
q_{DW}=\sum_{n=1}^{L_y/2} q_n.
\end{equation}

(b) When $L_y$ is odd the Hamiltonian can be decomposed into $\left\lfloor L_{y}/2\right\rfloor$ 4x4 blocks obeying the same symmorphic chiral symmetries. Each of these modes  carries charges $q_n=-1,0,1$ ($n=1,...\left\lfloor L_{y}/2\right\rfloor$). Additionally there exists one transverse mode $n=0$ which is similar as the one studied in the case of single-mode SSG chain. Thus, this transverse mode carries a charge
$q_0=(\pm 1 -\delta)/2$ in the case of solitons and $q_0=(\pm 1 + \delta)/2$ in the case of antisolitons. The total charge of the DW is
\begin{equation}
q_{DW}=\sum_{n=0}^{\left\lfloor L_{y}/2\right\rfloor} q_n.
\end{equation}

\end{document}